\documentclass{new_tlp}
 \usepackage{mathptmx}
 \usepackage{amsmath}
 \usepackage{color}
 \usepackage{amssymb}
 \usepackage[hyphens]{url}
 \usepackage[hidelinks]{hyperref}
 \newtheorem{definition}{Definition}
 \newtheorem{example}{Example}
 \newtheorem{proposition}{Proposition}
 \newtheorem{theorem}{Theorem}
 
 \newcommand{\comment}[1]{}
 \newcommand{\eat}[1]{}
 \usepackage{graphicx}
 \usepackage[T1]{fontenc}
 \usepackage{xspace}
 
 \def\mt{\tt}
 \def\1S{$\tt _{1S}$}
 \def\+1{{+}1}
 \newcommand{\bldl}{\smallskip\[\begin{array}{ll}}
 \newcommand{\cldl}{\[\begin{array}{ll}}
 \newcommand{\eldl}{\end{array}\]\rm}
 \newcommand{\prule}[2]{ \mt #1 \leftarrow & \mt #2 \\}
 
 \def\pfact#1#2{\mt  #1 &  \mt #2 \\}
 \def\pbody#1#2{ \mt #1 & \mt #2 \\}
 \def\inv{\vspace{-0.2cm}}
 \def\sinv{\vspace{-0.1cm}}
 
 \def\magg#1{ min \langle #1 \rangle}

 \def\prem{{\sc PreM}}
 
 \newcommand{\deals}{DeALS\xspace}
 \newcommand{\stitle}[1]{\vspace{0.6ex}\noindent{\bf #1}}
 
 \newcommand{\cfig}{Figure~}
 \newcommand{\cexp}{Example~}
\begin{document}

\title[Fixpoint Semantics and Optimization
of Recursive Datalog Programs with Aggregates]{Fixpoint Semantics and Optimization of\\Recursive Datalog Programs with Aggregates}


\author[Zaniolo et Al.]{
Carlo Zaniolo$^1$,$\;$Mohan Yang$^1$,$\;$Matteo Interlandi$^2\thanks{Work done while at UCLA.}$,$\;$Ariyam Das$^1$,$\;$Alex Shkapsky$^1$,$\;$Tyson Condie$^1$\\
$^1$University of California, Los Angeles\vspace{-0.2ex}\\
\email{\small  \{zaniolo,yang,ariyam,shkapsky,tcondie\}@cs.ucla.edu}\vspace{-2ex}\\\\
$^2$Microsoft\vspace{-0.2ex}\\
\email{\small  matteo.interlandi@microsoft.com}
}

\maketitle

\begin{abstract}
A very desirable Datalog extension investigated by many researchers in the last thirty years consists in allowing the use of the basic SQL aggregates \textbf{min}, \textbf{max}, \textbf{count} and \textbf{sum} in recursive rules. In this paper, we propose a simple comprehensive solution that extends the declarative least-fixpoint semantics of Horn Clauses, along with the optimization techniques used  in the  bottom-up implementation approach adopted by many Datalog systems. We start by identifying a large class of programs of great practical interest in which the use of \textbf{min} or \textbf{max} in recursive rules does not compromise the declarative fixpoint semantics of the programs using those rules. Then, we revisit the monotonic versions of \textbf{count} and \textbf{sum} aggregates proposed in~\cite{mazuran2012extending} and named, respectively, \textbf{mcount} and \textbf{msum}. Since mcount, and also msum on positive numbers, are monotonic in the lattice of set-containment, they  preserve the fixpoint semantics of Horn Clauses. However, in many applications of practical interest, their use can lead to inefficiencies, that can be eliminated by combining them with max, whereby mcount and msum become the standard count and sum. Therefore,  the  semantics  and optimization techniques of Datalog are  extended  to recursive programs with min, max, count and sum, making possible the advanced applications of superior performance and   scalability  demonstrated by BigDatalog \cite{shkapsky2016big} 
and Datalog-MC~\cite{yang2016scaling}. This paper is under consideration for acceptance in TPLP.

\end{abstract}

\begin{keywords}
Datalog, Constraints, Recursion, Aggregates.
\end{keywords}
\inv
\section{Introduction}
A growing body of research on  scalable data analytics has  brought a renaissance of interest
in Datalog because of its ability to specify, declaratively, advanced data-intensive applications
that execute efficiently over different systems and architectures, including
massively parallel ones~\cite{seo2013distributed,DBLP:journals/pvldb/ShkapskyZZ13,DBLP:conf/bigdataconf/YangZ14,aref2015design,wang2015asynchronous,yang2015parallel,shkapsky2016big,yang2016scaling}.
A common thread in this new generation of
Datalog systems is the use of aggregates in recursion, since aggregates  enable
the concise expression and efficient
support of much more powerful algorithms than those expressible by programs that are
stratified w.r.t. negation and aggregates.
However, the non-monotonic nature of common aggregates creates major issues,  and
extending the declarative semantics of Datalog
to allow aggregates in recursion
represents a difficult problem since the early days of Datalog~\cite{DBLP:conf/slp/KempS91,DBLP:conf/pods/GrecoZG92,ross1992monotonic}.
These approaches were seeking to achieve both (i) a formal declarative
semantics for deterministic queries using the basic SQL aggregates, min, max,
count and sum, in recursion
and  (ii) their efficient  implementation by extending techniques  of the early  Datalog 
systems \cite{DBLP:conf/iclp/MorrisUG86,DBLP:journals/debu/ChimentiOKTWZ87,DBLP:conf/vldb/RamakrishnanSS92,DBLP:journals/vldb/VaghaniRKSSLH94,ldl++}. 
As discussed in the related work section, while many of these  approaches only dealt with
some of the four basic aggregates, the proposal presented in \cite{ross1992monotonic}
covered all four. However it used
different lattices for different aggregates and thus incurred  in the problems 
described in \cite{van1993foundations}.  Therefore, the search for general solutions has motivated
several approaches that have addressed the problem using powerful semantics, 
such as answer-set semantics, that however require higher levels of computational 
complexity 
\cite{DBLP:journals/tplp/PelovDB07,DBLP:journals/tplp/SonP07,Swift2010,DBLP:journals/ai/FaberPL11,DBLP:journals/corr/GelfondZ14}.

Recently, the explosion of activity  on Big Data has produced a revival
of  interest in Datalog as a parallelizable language
 for  expressing  and  supporting efficiently Big Data Analytics
 ~\cite{seo2013distributed,DBLP:journals/pvldb/ShkapskyZZ13,wang2015asynchronous}. 
The algorithms discussed in those papers are 
graph algorithms or other algorithms that use aggregates in recursion for Big Data applications where performance is of the essence.
Thus, a convergence of declarative semantics and
 efficient implementation is critically needed and has
 motivated the research presented in this paper,
where we  propose  a  novel approach based  on 
 the concept of \emph{pre-mappability of constraints}.
Indeed, pre-mappability 
of constraints provides a simple criterion that (i)  the system optimizer
can utilize to push constraints into recursion, and (ii) the user
can  utilize to write  programs that use  aggregates in recursion,
with the guarantee that they have indeed a formal fixpoint semantics.

The rest of this paper is organized as follows. In the next section,
we introduce the problem of pushing constraints into recursion, then, in  Section~\ref{sec:fixpoint},
we present the rather surprising result that
in many programs of practical interest constraints applied to recursive  rules do not invalidate the formal
 fixpoint semantics of those rules. We thus introduce general sufficient
conditions under which we can push comparison constraints, extrema constraints, and
combinations of the two. 
With the treatment of
min and  max  in recursive rules completed, we move to the problem of
supporting  count and sum. In Section~\ref{sec:aggregates},
we present an in-depth treatment of this important subject by combining
the monotonic versions of these aggregates proposed in \cite{mazuran2013declarative} with the treatment of extrema proposed in the previous sections.
Seminaive evaluation in  the presence of constraint pushing is discussed in Section~\ref{sec:seminaive}.
Related work and conclusions presented in Sections \ref{sec:related} and \ref{sec:conclusion}
bring the paper to a closing.
\inv\inv\section{Comparison and Extrema  Constraints}\label{sec:extrema}
We will now propose motivating examples that also illustrate the Datalog
syntax and terminology we use, which are widely used in
the literature \cite{DBLP:books/mk/ZanioloCFSSZ97,shkapsky2015optimizing}.
  \begin{example}[Limited length paths from node a in a directed graph]
  \vspace{-2ex}
 \label{ex:llimited}\inv
 \bldl
\prule{(r_1)~~path(Y, Dy)} {arc(a, Y, Dy), Dy\geq0.}
\prule{(r_2)~~path(Y, Dy)}{path(X,  Dx) , arc(X, Y, Dxy), Dxy\geq 0, Dy=Dx+Dxy.}
\prule{(r_3)~~llpath(Y, Dy) } {path(Y, Dy),  Dy<143.~~~~~~~~~~~~~~~~~~}
\eldl
\end{example}
In the program of Example 1, 
 $\tt path(Y, Dy)$ is the \emph{head}  of $\tt r_1$, 
while $\tt arc(a, Y, Dy)$ and $\tt Dy\geq 0$ are the rule's \emph{goals}, where $\tt Dy\geq 0$ specifies a  \emph{lower-bound constraint} (symmetrically, $\tt Dy <143$ in $\tt r_3$ is an \emph{upper-bound constraint}).
In rule $\tt r_2$, the two-argument $\tt path$ is both the head and  a goal: thus $\tt r_2$ is a
\emph{recursive rule} and $\tt path$ is a recursive predicate. On the other hand,  non-recursive rule $\tt r_1$ with $\tt path$ as its head will be 
called an {\em exit} rule.
The arithmetic predicates used in $\tt r_2$
can be defined using Horn Clauses on Datalog$_{1s}$ \cite{DBLP:books/mk/ZanioloCFSSZ97}, and therefore 
our program has formal semantics based on the equivalence
of its model-theoretic, proof-theoretic, and least-fixpoint semantics~\cite{DBLP:journals/jacm/EmdenK76}.
The least-fixpoint of
a monotonic program  can be computed by iterating over the \emph{Immediate Consequence Operator} (ICO)
defined by the rules of the program, and Datalog researchers  have introduced techniques such as 
seminaive-fixpoint and magic-set to optimize this botton-up iterative computation into efficient implementations \cite{DBLP:books/mk/ZanioloCFSSZ97}. Even more significant
 optimization opportunities occur in the presence of constraints.
For instance, an unoptimized Datalog  computation for this program would apply  the upper-bound constraint $\tt Dy<143$ specified by the final rule $\tt r_3$
to all the  $\tt  path$  atoms generated by  iterating over rules  $\tt r_1$ and  $\tt  r_2$. Thus this constraint
is applied as a postcondition after all pairs have been generated, and does not improve the
efficiency of the computation; in fact this might never terminate  when our graph contains cycles.

 A very simple solution for this problem is to push the upper-bound constraint $\tt Dy <143$ into the
 recursive computation. We can hence rewrite the above program into the following, where the constraint has been transferred to the rules defining $\tt path$.
\begin{example}[Rewritten version of Example 1 where the constraint is transferred] 
 \label{ex:optimized_llimited}
   \vspace{-1ex}\inv
\inv \bldl
\prule{(r'_1)~~path(Y, Dy)} {arc(a, Y, Dy), Dy\geq0, Dy<143.}
\prule{(r'_2)~~path(Y, Dy)}{path(X,  Dx) , arc(X, Y, Dxy), Dxy\geq0, Dy=Dx+Dxy, Dy<143.}
\prule{(r'_3)~~llpath(Y, Dy) } {path(Y, Dy).}
\eldl
\end{example}
Thus we obtain a program where the constraint $\tt Dy<143$ has been \emph{transferred from
the final rule} $\tt r_3$ into the recursive program and applied to rules $\tt r_2$ and $\tt r_1$, i.e., {\em pre-mapped} to those rules.
Observe that the rules produced by this transformation have been renamed $\tt r'_3$, $\tt r'_2$ and $\tt r'_1$.  
The program so obtained provides
an optimized version of the original program that is equivalent to it inasmuch as the two 
programs deliver the same $\tt llpath$ results. This equivalence follows from the fact 
that  lengths of $\tt arc$ are $\geq 0$, whereby 
$\tt Dy<143$ in the body of $\tt r'_2$ 
implies that  $\tt Dy<143$  for $\tt Dy$ in the head of the same rule---i.e., pruned values are those 
that would be eliminated by the final ``$\tt < 143$'' constraint in the original $\tt r_3$ rule.

An even more significant optimization is possible when \emph{extrema constraints} (i.e., max and min aggregates)
are applied  to the results produced by recursive rules. For instance, say
that we substitute rule $\tt r_3$ in Example 1 with the following rule that uses 
the extrema constraint $\tt is\_min$. 
\vspace{-1.5ex}
\bldl
\prule{(r_4)~~spath(Y, Dy) } {path(Y, Dy),  is\_min((Y), (Dy)).}
\eldl
\vspace{-3.2ex}

\noindent The construct $\tt is\_min$ identifies 
 $\tt Dy$ as the \emph{cost argument} and  $\tt Y$ as  the \emph{group-by variable(s)}, 
using SQL terminology. 
Thus, $\tt is\_min((Y), (Dy))$ denotes that we want the min values of $\tt Dy$ for each
unique value of  $\tt Y$. 
The special notation used for min  warns the compiler that  its specialized internal
implementation should be used in its evaluation, along with the  query transformation 
and optimization techniques discussed next. However the semantics of rules with our
$\tt is\_min$ construct can be  reduced to that of rules with standard (closed-world)  negation, 
whereby the semantics of
 $\tt r_4$ is defined by the following two rules where $\tt is\_min$ is replaced by the negation that a lesser value exists:
 \footnote{This rewriting assumes that there is only 
 one $\tt is\_min$ in our program. In the presence of multiple occurrences, we will need to add a subscript to  
keep them distinct.}
\inv
\bldl
\prule{spath(Y, Dy) } {path(Y, Dy),  \neg lesser(Y, Dy).}
\prule{lesser(Y,Dy)}{path(Y, Dy), path(Y, Dy1), Dy1 <Dy.}
\eldl
Expressing $\tt is\_min$ via negation reveals
 the non-monotonic nature of extrema constraints,  whereby
their usage in the definition of recursive predicates 
can compromise the declarative
fixpoint semantics of the program.
\emph{Stratification}, where each predicate in 
the programs must belong to strata that are above 
those of their negated goals, can be used to avoid the semantic problems caused by using aggregates
in recursion. For instance, in our example  $\tt spath$ belongs to
a stratum that is above that of $\tt path$, whereby our program is assured to have
 a \emph{perfect-model semantics}~\cite{DBLP:conf/iclp/Przymusinski88}.
The perfect model of  a stratified program is a minimal model which 
is computed using an {\em  iterated fixpoint computation},  starting at the bottom stratum and moving
up to higher strata. In our example, therefore, all the possible paths will be computed
using rules $\tt r_1$ and $\tt r_2$,
before selecting values that are minimal using $\tt r_4$. This approach,
used by current Datalog compilers, can be very inefficient or
even non-terminating when the original graph 
of Example 1 contains cycles.
The naive re-writing similar to the one used to transfer the comparison constraint into 
our recursive rules  yields:

\begin{example}[Shortest path  from node $\tt a$]
\label{ex:shortestpath1}
  \vspace{-2ex}\inv
\cldl
\hspace{-1ex}\prule{(r''_1)~~path(Y, Dy)} {arc(a, Y, Dy), Dy\geq0, is\_min((Y), (Dy)).}
\hspace{-1ex}\prule{(r''_2)~~path(Y, Dy)}{path(X,  Dx) , arc(X, Y, Dxy), Dxy\geq0,Dy=Dx+Dxy, is\_min((Y), (Dy)) .}
\hspace{-1ex}\prule{(r''_4)~~spath(Y, Dy) } {path(Y, Dy).}
\eldl
\end{example}

However, while in Example \ref{ex:llimited} the  transferring of comparisons
into the recursive program produces the positive program of Example \ref{ex:optimized_llimited} 
which has a well-defined formal semantics, the program in Example \ref{ex:shortestpath1} above 
uses non-monotonic aggregates (alias negation if we replace them
with their  definitions) in recursive rules.  Thus, there is no guarantee that 
a formal semantics exists for the rewritten program, and the question  of whether this
is a correct optimization becomes meaningless.  Thus 
in  this  paper, we  provide a formal  minimal-fixpoint semantics 
 that holds for  Example~\ref{ex:shortestpath1},
and  for  large classes of programs of major practical 
interest ~\cite{ourpaper2}.

\inv\inv\section{Fixpoint and Constraints}
\label{sec:fixpoint}
The declarative  and constructive semantics of a Datalog program $P$ is defined in terms of the Immediate Consequences Operator (ICO) for $P$, denoted by $T_P(I)$, where $I$ is any Herbrand interpretation of $P$.
For positive Datalog programs, i.e., those without negation or aggregates, $T_P(I)$
is a monotonic continuous  mapping in the lattice of set-containment to which
the interpretation $I$ belongs, whereby we have the following well-known properties:
\begin{enumerate}

\item A unique minimal (w.r.t. set-containment) solution of the equation $I{=}T_P(I)$
always exists and it is known as  the {\em least-fixpoint} of $T_P$, denoted $\textit{lfp}(T_P)$;
$\textit{lfp}(T_P)$ {\em defines the formal declarative semantics of $P$}.

\item
For an immediate consequence operator $T$, with $\omega$ being the first infinite ordinal,
$ T^{\uparrow \omega}(\emptyset)$ is defined by letting  $ T^{\uparrow 0}(\emptyset)= \emptyset$, and
 $ T^{\uparrow n+1}(\emptyset)= T(T^{\uparrow n }(\emptyset))$. Then $ T^{\uparrow \omega}(\emptyset)$  denotes
 the union of $T^{\uparrow n}(\emptyset)$ for every  $n$.
The {\em fixpoint iteration} $ T^{\uparrow \omega}_{P}(\emptyset)$
defines the {\it operational semantics} of our program.
In most applications of practical interest,
the iteration converges to the final value in
a finite number of steps, and it can be stopped at the first
 integer $n{+}1$  where,
 $ T^{\uparrow n+1}_{P} (\emptyset)= T^{\uparrow n}_{P}(\emptyset)$.
  \end{enumerate}

\noindent For  positive Datalog programs we have that the operational and declarative semantics coincide.
 
In the previous section, we have discussed extrema and lower/upper-bound constraints which can be 
expressed and enforced upon rules by the addition of goals to the rules. We will now define conditions under which post-constraints on the results of recursive-rule computation can
be pre-mapped and applied to those rules.
Thus we introduce the notion of a constraint $\gamma$ being 
\emph{\it pre-mappable} to one or more rules, in which
case we will simply write that $\gamma$ is {\prem} to those rules. The {\prem} property will 
allow us to extend the 
fixpoint semantics to programs obtained by transferring a \emph{constraint} $\gamma$  from the final rule to  
the rules defining a recursive predicate\footnote{Note that here and in the following definition the $\gamma$ symbol is overloaded because it is used to express a constraint both as a literal in a final rule, and a pruning condition over instances.}.
Notions related to \prem~ are already used widely in optimizing
distributed computations.  For instance when a
set of tuples  is partitioned into $K$ subsets $S_1, \ldots, S_k$  we have that:
\sinv$$ {\tt OP} (\bigcup_{1\leq j \leq K} {\Large S_j}) = {\tt OP} (\bigcup_{1\leq j \leq K} {\tt OP} ({\Large S_j} ))$$
 Since this equality holds for {\tt OP} denoting the aggregates sum, min  and max, we will say that
these aggregates are \prem~\! to union. This property is frequently called preaggregation 
and exploited by the parallel execution  of $K$ {\tt OP} aggregates  on different nodes followed by a final execution of  {\tt OP}
on the results returned  by those $K$ nodes. The above property also holds 
when {\tt OP}  denotes an upper-bound or a lower-bound constraint;  however in that case,  the final  application 
of {\tt OP} can be omitted  given that upper/lower bound constraints distribute over union.  

\vspace{-0.5ex}\begin{definition}[The \prem~ Property] In a given Datalog program
let $P$ be the set of its  rules  defining either a recursive predicate or a set 
of mutually recursive predicates. Also let $T$ be the ICO defined by $P$.
Then, the constraint $\gamma$ will be said to 
be {\prem} to
$T$ (and to $P$) when, for every interpretation $I$ of
 $P$, we have that: $\gamma(T(I)) = \gamma (T(\gamma (I)))$.
\end{definition}
%

The importance of this property follows from the fact that if
$I = T(I)$ is a fixpoint for $T$, then we also have that
$\gamma(I) = \gamma(T(I))$, and when $\gamma$ is 
\prem~ to
the rules defining $T$ we also have that:\inv 
$$\gamma(I) = \gamma(T(I)) = \gamma(T(\gamma(I)))$$
Now,  let $T_\gamma$  denote the application of
$T$ followed by $\gamma$,  i.e., $T_\gamma(I)=\gamma(T(I))$, where $T_\gamma$ will be called
the \emph{$\gamma$-constrained}  ICO  for the $P$ rules. Now,  if
 $I$ is a fixpoint for $T$ and $I'= \gamma(I)$, then the above equality can be rewritten as:
 \vspace{-0.5ex}
$$I'=\gamma(I) =  \gamma(T(\gamma(I)))= T_\gamma(I')\vspace{-0.5ex}$$
Thus, when $\gamma$ is {\prem}, the  fact that  $I$ is a fixpoint for  $T$ implies
that $I'=\gamma(I)$  is a fixpoint for  $T_\gamma(I)$.
We will next focus on  cases of practical interest where the transfer of constraints 
under {\prem} produces optimized programs that are safe and terminating 
even when the original programs were not, and prove that the transformation is indeed
equivalence-preserving. Thus we focus  on situations where 
$T_\gamma^{\uparrow n}(\emptyset)= T_\gamma^{\uparrow n+1}(\emptyset)$, i.e., 
the fixpoint iteration converges after a finite number of steps $n$. 
The rules defining a  recursive predicate $p$ are those having as head $p$ or 
predicates that are mutually recursive with $p$.

\begin{theorem}
\label{thm:1}
In a Datalog program, let  $T$ be the ICO for the positive rules defining a recursive
predicate. 
If the constraint $\gamma$ is {\prem} to $T$,  and a fixpoint exists such that
$T^{\uparrow n+1}(\emptyset) = T^{\uparrow n}(\emptyset)$, for some integer $n$ then:
\inv\inv $$\gamma(T^{\uparrow \omega}(\emptyset)) = T_\gamma^{\uparrow n}(\emptyset)$$
\end{theorem}\inv
\begin{proof}
\sinv\sinv\inv ~~~~~~~~~~Let us assume that $T_\gamma^{\uparrow n+1}(\emptyset)= T_\gamma^{\uparrow n}(\emptyset)$ and  prove  by induction that  $\gamma(T^{\uparrow k}(\emptyset)) = T_\gamma^{\uparrow k}(\emptyset)$ 
holds for every non-negative integer $k$. Indeed, the  property holds for $k=0$ since $\gamma(T^{\uparrow 0}(\emptyset)) = T^{\uparrow 0}_\gamma(\emptyset) = \emptyset$.
 ~Moreover, since $\gamma$ is ~{\prem} ~ to ~$T$,  ~we have that the ~following ~equalities ~hold:
 $\gamma(T^{\uparrow k+1}(\emptyset)) = 
 \gamma(T (T^{\uparrow k}(\emptyset)))= 
 \gamma(T(\gamma(T^{\uparrow k}(\emptyset)))) = 
  \gamma(T(T_\gamma^{\uparrow k}(\emptyset)))= T_\gamma^{\uparrow k+1}(\emptyset).$ 


 
 Now we have that $\gamma(T^{\uparrow \omega}(\emptyset)) = \gamma(\bigcup_{k \geq 0}T^{\uparrow k}(\emptyset)).$ But  $\bigcup_{k \geq 0}T^{\uparrow k}(\emptyset)=
\bigcup_{m\geq n}(T^{\uparrow m}(\emptyset))$. In fact, the monotonicity of $T$
implies that for each $T^{\uparrow m}(\emptyset)$  with $m \leq k$ there is
a $T^{\uparrow m'}(\emptyset)$ with $m' > k$ containing it. 
Now, our $\gamma$ constraints are also  {\prem} to union, and thus we have that
 $\gamma(\bigcup_{m\geq n}T^{\uparrow m}(\emptyset))= 
 \gamma(\bigcup_{m\geq n}\gamma(T^{\uparrow m}(\emptyset))) = 
 \gamma(\bigcup_{m\geq n}(T_\gamma^{\uparrow m}(\emptyset)))= 
 \gamma(T_\gamma^{\uparrow n}(\emptyset))$  
 since, for $m\geq n$ all the $T_\gamma^{\uparrow m}(\emptyset)$ are identical.
Finally, we have that $\gamma(T_\gamma^{\uparrow n}(\emptyset)) = \gamma(\gamma(T^{\uparrow n}(\emptyset))) = T_\gamma^{\uparrow n}(\emptyset)$ because applying twice the same constraint is equivalent to a single application.
 Thus we conclude that $\gamma(T^{\uparrow \omega}(\emptyset)) = \gamma(T_\gamma^{\uparrow n}(\emptyset)) =
 T_\gamma^{\uparrow n}(\emptyset)$. 
\end{proof}

\sinv We can next prove that the fixpoint \ $T_\gamma^{\uparrow n}(\emptyset)$  so obtained is in fact 
a minimal fixpoint for $T_\gamma$.  

\begin{theorem}
If $\gamma$ is {\prem} to $T$ and  $T_\gamma^{\uparrow n}(\emptyset) = T_\gamma^{\uparrow n+1}\!(\emptyset)= I$,
then $I$ is  a minimal fixpoint for $T_\gamma$.
\end{theorem}
\inv\begin{proof}\sinv\sinv\inv ~~~~~~~~~~
 If  $I' \subseteq T_\gamma^{\uparrow n}(\emptyset)$
is a fixpoint, then  $I' = T_\gamma^{\uparrow k}(I')$ for every integer $k$. Now,
starting with  $ T_\gamma ^{\uparrow  0}(I') = I' =\gamma ( T ^{\uparrow  0 }(I'))$ we  can apply 
the  inductive reasoning process  used in Theorem 1 to conclude 
that  the equality $ T_\gamma^{\uparrow  k}(I') = \gamma ( T ^{\uparrow  k}(I'))$ holds for each $k$.
Therefore the union, for all $k$,  of  the sets on the left side of the equation is equal 
to the union of all the sets on the right side.  But  the unioned sets on the left are equal to $I'$,
whereas all those on the right side define $\gamma ( T ^{\uparrow  \omega}(\emptyset))$, since $\emptyset \subseteq I^\prime$.
But Theorem 1 has shown that $ \gamma ( T ^{\uparrow  \omega} (\emptyset)) = T_\gamma^{\uparrow n}(\emptyset)= I$.
Thus 
no proper subset of $ \gamma ( T ^{\uparrow  \omega} (\emptyset))$ can be a fixpoint for $T_\gamma$.
\end{proof}

Therefore, constraints which are {\prem} can be transferred into recursive rules and this represents a significant optimization
which, in many cases,  also
produces programs where the fixpoint  iteration 
terminates even when the iterated fixpoint computation on the original program did not. 
Moreover, the termination of the fixpoint iteration for the rewritten program ensures that we obtain a minimal fixpoint that is equivalent to the original program with respect to the predicates that were
constrained by the {\prem} constraint $\gamma$.
Thus the well-known compilation techniques of positive programs, including
magic set 
 and seminaive computation,
can be applied to programs optimized by the transfer of {\prem} constraints.

While the problem of determining general classes of  {\prem}
constraints leaves much room for further research, we have already determined
classes of {\prem} constraints that  hold for programs of great practical interest, such as those discussed next.
\inv
\section{Transferring {\large \prem} Constraints}
We will next establish basic sufficient conditions under which 
transferability of extrema and comparison constraints hold for programs such
as those in Examples 1 and 2.  In those examples we have a
final rule (in the next stratum) applying a constraint to some recursive predicate, which we call the {\it cost 
predicate}, defined by a set of rules which will
be called the {\it cost rules}.  
The key intuition behind {\prem} constraints is that some form of monotonicity must be preserved while constraints are applied in
recursion. A form of monotonicity preserved under extrema 
is one where the $\geq$  or $\leq$  relationships 
holding for cost variables in the bodies of two instances of a rule 
implies that the same relationship holds for the cost variables in their heads.
A stricter condition is the one in which the cost argument in the head is required to be $\geq$ (resp. $\leq$) than the related cost argument in the body. When this condition is satisfied, upper (lower) bound constraints are {\prem} (Proposition 2), because  once the constraint is not satisfied for a certain tuple combination and a cost argument, it cannot be satisfied for increased (decreased) costs.
Next, we will make the above intuitive explanation more formal.
For simplicity, we assume that  there is  no predicate
that is mutually recursive with our cost predicate.
The cost rules  have two types of goals. The  
goals of the first type use {\it interpreted} predicates, such 
as arithmetic comparisons and other built-ins
whose semantics and implementation are pre-defined in the system. The 
remaining goals will be called {\it  regular goals}  which refer to
regular predicates, i.e.  predicates whose  meaning is  being defined by
the program at hand.  The cost predicates appearing in the head of our rule are 
also regular predicates.
We will use the notation $rg(r)$ to denote the 
conjunction of the regular goals in the body of a rule $r$.
Thus in our Example 1, $rg({\tt r_2})= "{\tt path(X, Dx), arc(X, Y, Dxy)}"$.

Each constraint in the final rule defines
a cost argument for a particular regular predicate:  e.g., 
in our previous examples both the constraints >143 and {\tt is\_min} established 
the second argument of $\tt path$ to be the cost argument.
For simplicity of discussion, we will only consider cases where
for each regular predicate $p$ there is only one cost argument
denoted $ca(p)$.
Finally, we assume that rules (and predicates) are \emph{instantiated} as usual by assigning to
their  variables values taken from a suitable universe (e.g., a Herbrand universe that also
includes numbers).  
Now,  we will define {\em valid instances of a rule} to be those where  variables are instantiated to
values that make all the interpreted goals true.  Thus for our rule $\tt r_2$ in Example 1, arbitrary 
values can be assigned to the variables $\tt X$ and $\tt Y$. However the values assigned to 
$\tt Dx, Dxy$  and $\tt Dy$ must satisfy the conditions $\tt Dxy \geq 0 \land Dy= Dx+Dxy$.
In order to determine if extrema are {\prem}, we  have to  provide a sufficient condition whereby the validity of the $\geq$  or $\leq$  relationships for  cost variables in the bodies 
of two instances of the given rule implies that the same relationship holds for the cost variables in their heads.
Therefore  we will  introduce the notion  that \emph{inflated} cost-predicates  
in the body of the rule also \emph{inflate} the  
and cost-predicates of the heads of the rules.


\inv\vspace{0.4ex}\begin{definition}[Inflating / Deflating cost predicates]
Let  $\textbf{p}^\prime$ and $\textbf{p}^{\prime\prime}$ 
be two instances of a cost predicate
$p$ where  non-cost arguments 
of $\textbf{p}^\prime$ are identical to those of $\textbf{p}^{\prime\prime}.$ Then when $ca(\textbf{p}^\prime) \geq ca(\textbf{p}^{\prime\prime})$ we say 
that $\textbf{p}^{\prime}$ \emph{inflates} $\textbf{p}^{\prime\prime}.$ Conversely, when $ca(\textbf{p}^{\prime}) \leq ca(\textbf{p}^{\prime\prime})$
 we say that  $\textbf{p}^{\prime}$ \emph{deflates} $\textbf{p}^{\prime\prime}.$ 
\end{definition}

\vspace{-0.7ex}
\sinv\begin{definition}[Inflating / Deflating conjuncts of regular goals]
Let $r$ be a rule and $\textbf{r}^{\prime}$ and $\textbf{r}^{\prime\prime}$ be two of its instantiations where all the
regular non-cost goals are identical and each regular cost goal
of $\textbf{r}^{\prime}$ in $rg(r)$ inflates (resp. deflates) the corresponding goal in 
$\textbf{r}^{\prime\prime}$. Then we say that the regular
goals of $\textbf{r}^{\prime}$ \emph{inflate} (resp. \emph{deflate}) the regular goals of $\textbf{r}^{\prime\prime}$.

 \end{definition}
 
\sinv For example, consider the following two instances of a rule $r$ which are identical
everywhere  but in their cost arguments; the interpreted goals use the square-root function $\tt sqrt$:
\inv
\bldl
\prule {\textbf{r}^{\prime} : path(y,  dy^\prime)}{path(x, dx^\prime), arc(x, y, dxy), 
dxy \geq 0,dy^\prime = dx^\prime+ sqrt(dxy).}
\prule {\textbf{r}^{\prime\prime} : path(y,dy^{\prime\prime})}{path(x,dx^{\prime\prime}),arc(x,y,dxy),dxy \geq 0,
dy^{\prime\prime} = dx^{\prime\prime}+sqrt(dxy).}
\eldl

\noindent If we assume that $\tt dx^{\prime\prime} > dx^\prime$, then  
${\tt path(x, dx^{\prime\prime})}$ represents an
arbitrary inflation of ${\tt path(x, dx^\prime)}$. 
Likewise, the  regular goal conjunction of the second rule,
i.e., $\tt path(x,dx^{\prime\prime}),arc(x,y,dxy)$, represents an arbitrary
inflation of regular goal conjunction of the first rule. 
We now have the following condition under which  extrema constraints become {\prem}.

\vspace{-0.5ex}\begin{definition}[Inflation / Deflation preserving  rules]
 A rule $r$ will be said to be \emph{inflation-preserving} (resp. \emph{deflation-preserving}),
when for every valid instance $\textbf{r}^{\prime}$  of $r$, 
and for every $V$ that is an 
inflation-preserving (resp. deflation-preserving) instance of the regular goals of $\textbf{r}^{\prime}$,  there 
exists a valid instance $\textbf{r}^{\prime\prime}$ of $r$  which (i)
has $V$ as its regular goals,  and (ii) its head that inflates (resp. deflates) the head of $\textbf{r}^{\prime}$.
\end{definition}

\noindent To apply this definition to our example we must check that  for each  $\tt path(x,dx^{\prime\prime}),arc(x,y,dxy)$
that inflates  $\tt path(x,dx^{\prime}),arc(x,y,dxy)$ 
there exists a valid instance of the interpreted goals
of $\textbf{r}^{\prime\prime}$
that delivers a head that inflates that of $\textbf{r}^\prime$. Now, $\tt sqrt$ 
has two valid interpretations, one that returns a positive square-root value and the other 
that returns its opposite negative value. Now, if we only allow the negative  interpretation of  $\tt sqrt$, we see that for
$\tt dy^{\prime\prime} \geq dy^\prime$,  
 $\textbf{r}^{\prime\prime}$ does not inflate  $\textbf{r}^\prime$, and thus $r$ is not inflation-preserving; however, if we only 
allow the positive interpretation {\em or we allow both}, then our rule becomes inflation preserving.

Next, let us call {\em procedure} the set of  rules in a program  defining the same predicate, while 
 those defining a cost predicate will be called  a cost procedure.
A  cost procedure will be called inflation-preserving (resp. deflation-preserving)
when all its rules are inflation-preserving  (resp. deflation-preserving). Moreover,   by 
applying a {\prem} constraint $\gamma$  to a  cost procedure we mean that 
the constraint is applied to each rule in the procedure. Then, we have:

\begin{proposition}
Min constraints are {\prem} to  deflation-preserving cost procedures, 
and max constraints are {\prem} to inflation-preserving cost procedures.
\end{proposition}
\inv\begin{proof}
\sinv\sinv\inv ~~~~~~~~~~
To  prove that $\gamma$ is {\prem} we need to prove that for each rule  $r$ of our cost procedure $\gamma(T(I))= \gamma(T(\gamma(I)))$.  If $r$ is non-recursive (i.e., an exit rule) the result is 
immediate since $\gamma(T(\emptyset))= \gamma(T(\gamma(\emptyset)))$. 
If $r$ is a recursive rule and $\textbf{r}'$ is an instance of $r$, then the application of $\gamma = \tt max$ to $I$  invalidates $\textbf{r}'$ iff
there is another instance of $r$ say $\textbf{r}''$ which inflates the regular goal conjunct
of $\textbf{r}'$. But since this rule is inflation-preserving the head of $\textbf{r}''$ inflates that
of $\textbf{r}'$ and it will thus be eliminated by the  application of the $\gamma = \tt max$ outside the parentheses. 
Therefore we conclude that $\tt max$ is 
{\prem} to inflation-preserving cost procedures, and symmetrically  
$\tt min$ is {\prem} to deflation-preserving cost procedures.
\end{proof}

\noindent For instance, returning to rules $\tt r_1$ and $\tt r_2$ in Example 1, we see that
they are both inflation-preserving and deflation-preserving. Thus, we can safely transfer either 
the $\tt is\_min$  or 
$\tt is\_max$ constraint from $\tt r_4$, as shown by 
Example 3 that was produced by transferring $\tt is\_min$.

We next provide transferability conditions for upper/lower-bound
constraints.

\begin{definition}[Ascending / Descending mappings from rule bodies to their heads]
A rule $r$ defines an \emph{ascending}  (resp. \emph{descending}) \emph{mapping} 
when in each of its valid instances $\textbf{r}$, the
cost argument in the head is $\geq$ (resp. $\leq$) the cost argument in each of the cost goals of $\textbf{r}$.
\end{definition}
\vspace{-0.5ex} Thus, for our previous rule using 
$\tt sqrt$, if we only use the
positive interpretation of  $\tt sqrt$,
then our rule defines an ascending mapping,
but  under  negative interpretation of $\tt sqrt$ we have 
a descending mapping since $\tt dy^{\prime} < dy$.  Thus if we allow both interpretations the rule neither  defines  an ascending mapping nor  a descending one. Likewise,
rules $\tt r_1$ and $\tt r_2$ in Example~1 define an ascending mapping, whereas they define a descending mapping if we assume that the arc lengths in our graphs are $\leq 0$. 
Note that inflation-/deflation-preserving differ from ascending/descending mappings in that the former does not require the cost in the head to be $\geq$ (resp. $\leq$) each cost argument in the body. For example, if in $\tt r_2$ we replace the goal $\tt Dy=Dx+Dxy$ with $\tt Dy=(Dx+Dxy) - 99$, the rule remains inflation-preserving, while it no longer defines an ascending mapping.

Now, with conditions such as $< 143$ and $\leq 143$ defining   \emph{upper-bound constraints} (and dually 
for \emph{lower-bound constraints}), 
we have that:

\sinv\begin{proposition}\label{thm:prop4}
Upper-bound  constraints are {\prem} to cost procedures
 defining ascending mappings, whereas lower-bound constraints
are {\prem} to cost procedures defining descending mappings.
\end{proposition}
\inv\sinv\begin{proof}\sinv\inv\inv ~~~~~~~~~~
Say that an  upper-bound constraint $\gamma$ has been
applied to a rule $r$ in our cost procedure  defining an ascending mapping.
The proof that $\gamma(T(I))= \gamma(T(\gamma(I)))$ is trivial for each exit
rule since they are applied upon $I= \emptyset$. Say now that $T$ is the ICO of a recursive rule $r$. Then 
the application of our {\prem}
 upper-bound constraint eliminates each instance $\textbf{r}'$ which 
fails the constraint; but  since $r$ defines an ascending mapping, the head  of $\textbf{r}'$
will then fail the same upper-bound constraint. Thus the 
 upper-bound constraints are {\prem} to all  rules of a cost procedure
defining an ascending mapping. Symmetrically for lower-bounds and descending mappings.
\end{proof}

We see that the cost procedure in Example \ref{ex:llimited} defines an ascending mapping  and thus 
the transfer of upper-bound  ${\tt Dy <143}$ done in Example 2 is correct. The rules composing the procedure also
define descending mappings and would also support the transfer of lower-bound constraints.
However, it is also very important to observe how upper-bound and lower-bound
constraints interact and often interfere with extrema.
For instance, say that rule $\tt r_4$ applied to  our Example 2
specified both the  constraints $\tt Dy<143$ and $\tt is\_min((Y), (Dy))$. Then we see that each of the two and their conjunct
{\prem} and can be pushed into the procedure. But in general,
the fact that each constraint is {\prem} does not assure that
its conjunct is too. For instance, after the transfer of $\tt Dy<143$ rule $\tt r'_2$  is still deflation-preserving but no longer inflation-preserving.
Thus, while 
$\tt is\_min((Y), (Dy))$ is still transferable after we transfer $\tt Dy<143$, the constraint $\tt is\_max((Y), (Dy))$
 is not. Symmetrically if we transfer the constraint $\tt Dy\geq 143$
 instead of $\tt Dy<143$, our rules become inflation-preserving, whereby max constraints can be transferred, but min constraints cannot since the rules are no longer deflation-preserving.

\inv\section{Monotonic Aggregates}
\label{sec:aggregates}

At the core of the approach proposed in \cite{mazuran2012extending}
there is the observation that the cumulative version of standard count is monotonic in the lattice of set containment. Thus \cite{mazuran2012extending} introduce 
 $\tt mcount$  as the 
aggregate function that returns all natural numbers up to the cardinality of the set.
The use of $\tt mcount$ in actual applications
is illustrated by the following example that is similar to one
proposed by \cite{ross1992monotonic}.

\vspace{-0.5ex}
\begin{example}[Join the party once you see that three of your friends have joined]
\label{ex:attend1}
The organizer of the party will attend, while other people
will attend if the number of their friends attending 
is  greater or equal to 3, i.e., ${\tt Nfx \geq 3}$.
\cldl
\prule{ attend(X)}{\hspace{-1.4cm}organizer(X).}
\prule{attend(X)}{\hspace{-1.4cm}cntfriends(X, Nfx), Nfx \geq 3.}
\prule{cntfriends(Y, N)}{\hspace{-0.2cm}attend(X), friend(Y,X), mcount((Y),(X), N ).}
\eldl
\inv
\end{example}
\inv

Here too we use  the body notation  for our   monotonic 
count-distinct  aggregate, $\tt mcount((Y), (X), N)$, where   
its first argument  $\tt (Y)$  is  the group-by argument; its second $\tt (X)$, is the 
\emph{count  argument}, and the third is
the result of the count aggregate. The count argument consists of one or more sub-arguments, whereas the 
group-by can consists of zero or more sub-arguments.
As proven in \cite{mazuran2012extending}, 
the semantics of  $\tt mcount((Y), (X), N)$
can be reduced to that of Horn Clauses.
The use of $\tt mcount$ in recursion
allows us to express programs, such as counting
paths between two nodes, that are not expressible using the standard aggregates
in stratified programs \cite{mumick1995expressive}.
Moreover, since $\tt mcount$ is monotonic, the fixpoint semantics  and optimization techniques of Datalog remain valid. In particular, we can optimize our program by pushing max into the recursion.
%

By  the query goal  $\tt attend(N)$ the
program in Example \ref{ex:attend1} returns the names of the people 
who attend the party. To find out the 
number of friends each attendee will have at the party,
we can use the query goal  $\tt  cntfriends(Y, X)$, which however
is likely to return many  answers, since
for an attendee  with $k$ friends it will return
his/her name $k$ times.  A solution to this problem  
is to use the following rule to return the final 
results~\footnote{Along with the people who count three
friends at the party, this will also return those who
count only one or two friends among those, who can
be removed by an additional final condition $\tt N \geq 3$.}:
\cldl
\prule{fcount(Y, N)} {cntfriends(Y, N), is\_max((Y), (N)).}
\eldl
The obvious question now is whether
$\tt is\_max$ can be transferred into the recursive rules
defining $\tt cntfriends(Y, N)$, which is true
if these rules are inflation-preserving. We have
previously examined  the case where there is only 
one recursive rule, and here we can reduce our
problem to that situation by taking
the composition of the two recursive rules in 
Example~\ref{ex:attend1}, obtaining the following rule:
\cldl
\prule{cntfriends(Y, N)}{cntfriends(X, Nfx), Nfx \geq 3, 
friend(Y, X), mcount((Y),(X), N).}
\eldl

Now, we  must determine if the condition of inflation
preservation is satisfied so that we can transfer the max into our
recursive rules. 
Since there is no valid instance of this rule for $\tt Nfx <3$, we only
need to consider instantiations in which  $\tt Nfx \geq 3$.
Note that the goal $\tt mcount((Y),(X), N)$ is 
independent from $\tt Nfx$, and the count can not decrease: thus, the rule is inflation-preserving.


\begin{example}[Join the party once three of your friends have joined using standard count]
\label{ex:attend5}\inv
\cldl
\prule{ attend(X)}{organizer(X).}
\prule{attend(X)}{cntfriends(X, Nfx), Nfx \geq 3.}
\prule{cntfriends(Y, N)}{attend(X), friend(Y,X),count((Y),(X), N ).}
\prule{fcount(Y, N)} {cntfriends(Y, N).}
\eldl
\inv\inv
\end{example}
This formulation using $\tt count$ is more efficient
than the one using  $\tt mcount$ since it implies that
condition $\tt Nfx \geq 3$ is no longer
checked for each +1 increment: it is instead checked 
for each new count value delivered
by the previous step in the iteration. At each iteration,
the old count value is incremented to include all the newly-found friends. 
Once the fixpoint is reached,  $\tt fcount(Y, N)$ returns, for 
each person  $\tt Y$  attending the party, the actual count $\tt N$ of
$\tt Y$'s friends at the party.


We might also allow users to write programs such as that 
in Example \ref{ex:attend5} where $\tt count$ is used instead of $\tt mcount$ in recursion. The compiler will process these programs by interpreting $\tt count$ as a max $\tt mcount$ and
transferring max out of recursion before checking that the resulting program is 
indeed inflation preserving. If this check succeeds then the original program
is executed, otherwise an error message, or a stern warning, is returned to the user.

\inv
\subsection{From Monotonic SUM to Regular SUM } 
The notion of monotonic sum $\tt msum$ for positive numbers
was also introduced in \cite{mazuran2012extending} as an extension of $\tt mcount$.  
For instance, say that we have a set of facts following the template $\tt part(Pno, Cost)$,
where for each part number we have its cost expressed as a positive integer.
To compute the total value of the parts on store, we can add up their costs using the sum
aggregate, as follows:
\inv
\cldl
\prule{total(T)} {part(Pno, C), sum((~),  (Pno, C), T).}
\eldl
Here $\tt (Pno, C)$ states that the sum is taken over
every  $\tt C$ appearing under
a distinct $\tt Pno$. As such, the semantics of sum is easy to reduce to that of count,
by  first expand each cost  $\tt C$ into $\tt C$ pairs $\tt (C, J)$, with $\tt 1\leq J\leq C$ 
and then compute the sum by counting these pairs. For instance, the pair
$\tt (bolt, 4)$  is represented by the four tuples: 
$\tt (bolt, 4, 1), (bolt, 4, 2), (bolt, 4, 3), (bolt, 4, 4)$. 
The meaning of $\tt sum((~),  (Pno, C), T)$ can now be defined by the pair of goals $\tt int\_up2(C, Int), count((~),  (Pno, C, Int), T)$, where $\tt int\_up2(C, Int)$ produces all the integers up to $\tt C$.
Thus our previous rule can be rewritten as:
\cldl
\prule{total(T)} {part(Pno, C), int\_upt2(C, Int), count((), (Pno, C, Int), T).}
\prule {int\_up2(C, 1) }{ C>0.}
\prule {int\_up2(C,J1) }{ int\_up2(C, J), J1=J+1, J1 \leq C.}
\eldl
In a totally analogous way we can define $\tt msum$ via $\tt int\_up2$ and $\tt mcount$, whereby, e.g., 
$\tt msum((), (C, Int), T)$ is defined as equivalent to $\tt int\_upt2(C, Int), mcount((), (C, Int), T)$.
Furthermore, the regular sum is simply the result of applying the max aggregate to $\tt msum$,
and this leads to many useful applications where sum is used directly in inflation-preserving recursive rules.
For instance say  that we have the classical part-explosion problem
where we have a set of basic parts with their cost $\tt basic(Part, Cost)$,
and then we have information about assembled parts, describing for
each assembled parts their subparts and quantities in the assemblage 
$\tt assb(Part, SubPart,$ $\tt Qty)$. To compute the cost of all parts we can write the following program:

\begin{example}[Part explosion problem: the cost of parts as sum of the cost of its subparts]\inv
\cldl
\prule{cost(Part, Cost)}{basic(Part, Cost).}
\prule{cost(Part, Ncost)}{assb(Part, SP, Qty),cost(SP, Cost),}
\pbody{}{CQ = Cost*Qty, sum((Part), (SP,CQ),Ncost).}
\prule{finalcost(Part, Cost)}{cost(Part, Cost).}
\eldl
\end{example}
Presented with this program, the compiler must make sure it can be viewed as the result of  transferring the max to the program obtained by replacing $\tt sum$
with $\tt msum$ in the second rule. To achieve this, the compiler will check that (i)
the arguments $\tt CQ$ of $\tt msum$ are all positive, and (ii) that the rules are inflation-preserving. 

To enable compile-time  checking of (i)
the user must add the goal $\tt CQ>0$ to
the recursive rule in our example. In the absence
of this, or equivalent conditions, the compiler
returns a stern warning message, but the user will still be 
able to proceed with execution under the optional
trust-but-verify policy that reports this violation
as a run-time error. This trust-but-verify option represents
a very useful debugging tool, and is 
also available if, for any reason, the compiler cannot
verify (ii).

This example illustrates the great benefits that should be expected from allowing sum in recursion,
for those situations where its formal semantics can be reduced to the transfer
of a final max into $\tt msum$.  However, there seems to be no obvious application that 
requires the use of $\tt msum$ instead of $\tt sum$. Moreover, say
that the regular sum of costs of our parts amounts to $50,000$:  the msum will return the first $50,000$ integers, thus causing serious inefficiencies.
Thus while the compiler might use the notion of $\tt msum$ to verify that the given program
using $\tt sum$ in its recursive rules has a  fixpoint semantics, $\tt msum$ should not be part of the actual language.  Since it is implemented directly without 
calling upon its equivalent count representation, sum  can also be applied to 
arbitrary positive numbers represented in scientific 
 (mantissa and exponent) notation, given that they can be viewed
 as large integers~\cite{mazuran2012extending}.

\sinv\inv\section{Seminaive Optimization}
\label{sec:seminaive}
So far,  we have focused on optimizing the naive fixpoint computation, which basically generates a new set of tuples from the whole set obtained in the last
step.  We can now perform the seminaive improvement which can be used to improve the computation of the program
obtained from the transferring of constraints.  For standard Datalog, the seminaive computation
operates as follows:
\begin{enumerate}
\item [A.] Keep track of the step at which each new atom was created,

\item [B.]  Modify the rules into their differential version, which exploits 
 the fact that rule instances whose bodies only contain atoms produced in older
steps cannot produce new atoms, and

\item [C.] Use  the modified rules to produce atoms, enforcing the no-duplicate
constraints as the atoms are produced.
\end{enumerate}

For programs obtained by the pushing of constraints, A and B  above remain
 the same, but C is significantly extended beyond deduplication  to enforce
constraints, whereby new atoms produced in B
might not be kept and existing atoms might instead
be eliminated. For constraints involving extrema,
for instance, besides the removal of duplicate atoms, step C will
also make sure no atom that violates the extremum constraint
is added to the working set.
For instance,  if a program with max constraints produces a new
$\tt (b, D)$, then this new pair is added to the working set only
if this does not
already contain a pair $\tt (b, D1)$ with $\tt D1 \geq D$, and
if this new pair is inserted  then each pair 
 $\tt (b, D2) $ with  $\tt D2 < D$ is deleted from the
 working set. Symmetric properties hold for min aggregates.
 The basic A, B, and C steps are also used to compute $\tt mcount$  with the proviso that
 every non-duplicate atom produced in C will increment the running count by one, and the new count value is added to the output. 
For count and sum, the 
 running counts and sum are incremented at step C, but atoms are
 only produced at the end of the fixpoint computation. Further 
 refinements of this strategies support the
 optimizations of \cite{shkapsky2016big}.

\eat{Thus,  the management of step C is quite obvious  for
extrema (and comparison-with-constant) constraints; it might be
less obvious in other cases, such as the coalesce program in
 Example \ref{ex:coalescing} in the Appendix, where we coalesce two tuples T1 and T2
 that overlap, and obtain  a tuple containing both T1 and T2. Thus the tuples
 that in step B were recognized as new-tuple generators are  exactly
 those that must be eliminated in our step C, a property that allows us
 to reduce the cost of our computation.\inv
 }
\inv\inv\section{Related Work}
\label{sec:related}
Several solutions, including  \cite{Kemp89propagatingconstraints,Srivastava92pushingconstraint}
addressed  the problem of
pushing comparison constraints into recursive rules, without exploring
 relationships with the problems of supporting extrema and other
aggregates into recursion.
Supporting aggregates in recursion is a difficult problem which
has been the topic of much previous research work. Several previous approaches
focused primarily on providing a formal semantics that could accommodate
the non-monotonic nature of the aggregates. In particular~\cite{DBLP:conf/vldb/MumickPR90}
discussed programs that are stratified w.r.t. aggregate operators
and proved that a perfect model exists for these programs. Then,~\cite{DBLP:conf/slp/KempS91} defined extensions of the well-founded semantics
 to programs with aggregates, and later showed that
 these might have multiple and counter-intuitive stable models.
The notion of cost-monotonic  extrema aggregates was introduced by~\cite{DBLP:journals/jcss/GangulyGZ95},
 using perfect models and well-founded
semantics, whereas~\cite{DBLP:conf/pods/GrecoZG92} showed that
to express greedy algorithms they require the  don't-care
 non-determinism of the stable-model semantics
provided by the choice construct.

The problem of  optimizing  programs with extrema by
early pruning of non-relevant facts
was studied in  \cite{DBLP:conf/vldb/SudarshanR91}.
A more recent work~\cite{Furfaro:2002:PEA:599471.599473} devises an algorithm for pushing max and min constraints into recursion while preserving query equivalence under certain specific monotonicity assumptions.
Those approaches provided effective optimization techniques for 
 min and max aggregates in recursion, but did not
consider 
count and sum aggregates. The idea of relating comparison with extrema constraints is also used in~\cite{Liu:2012:CED:2398857.2384645} to optimize logical quantification.

A general approach to deal with all four aggregates,
 was  proposed by~\cite{ross1992monotonic} who advocated the use of
 semantics based on specialized lattices, different from
 set-containment, whereby each aggregate will then define a  monotonic
 mapping in its specialized lattice.
However,~\cite{van1993foundations} pointed out that automatically determining the correct lattices would be difficult in practice,
and this was one of the causes that
prevented the deployment of the monotonic-aggregate idea in query languages for the
following twenty years.
Interestingly, a similar idea (with similar limitations) was more recently proposed in ~\cite{Swift2010} and in several other works e.g.,~\cite{Zhou:2010:MTD:1916727.1916958,zhou2015planning} in the context of tabling in Prolog. Also here is programmer's 
 duty to make sure that a proper partial order / lattice type is used for the program at hand. 

A renewed interest in Big Data analytics brought a revival of  Datalog as a parallelizable language
 for  expressing more powerful  graph and data-intensive algorithms---including many that require aggregates in recursion~\cite{seo2013distributed,DBLP:journals/pvldb/ShkapskyZZ13,wang2015asynchronous}.
The solution proposed here (and previously sketched in \cite{DBLP:conf/amw/ZanioloYDI16}) builds on the monotonic count and sum
proposed in \cite{mazuran2013declarative} and provides the foundation of
the efficient and scalable systems discussed in 
~\cite{shkapsky2015optimizing,yang2015parallel,shkapsky2016big,yang2016scaling}.

\vspace{0.5ex}\inv\inv\section{Conclusion}
\label{sec:conclusion}
In this paper, we have presented a unifying solution to the  problem of allowing the use of aggregates
in recursive Datalog programs.  The widely recognized practical importance of  solving this problem
has motivated  in-depth  formal investigations by many excellent researchers who during the  last 25 years studied and
proposed specialized solutions for  parts of this multi-faceted problem.
The main advantage of the unified treatment for the four basic aggregates, min, max, count, and sum,
proposed in this paper is  that it allows a simple specification of  algorithms  of great practical
interest with the guarantee that they have a minimal model semantics that is computed 
via the standard  fixpoint iteration, which can be  further refined by the standard optimization
techniques of bottom-up  Datalog implementations,
such as seminaive fixpoint computation, magic-set, left-linear and right-linear recursive rules.  
The practical significance  of these results are  demonstrated by 
the fact that, using these
techniques,  our BigDatalog system for Apache Spark was able to outperform graph applications
written in GraphX and running on the same system~\cite{shkapsky2016big}, suggesting that Datalog,
as a general-purpose high-level language, might be able to outperform high-level languages in the 
specialized application domains they were designed for.

Practical experience  with real-world
applications \cite{ourpaper2} also taught us  important lessons
on the dual role of fixpoint semantics and optimization that
 inspired the title and the research of this paper.
Initially, we followed the query
optimization approach since it appeals to intuition  and was
used by many researchers in the past. This
led to the equivalent minimal-fixpoint semantics for programs that,  via {\prem},
can exploit extrema and other constraints in recursive rules.
However, after  developing   applications in several  domains
\cite{ourpaper2}, we realized that specifying {\prem} constraints
in recursive rules represents a natural way to write those applications
(perhaps because  of the similarity with their iterative
procedural counterparts we are familiar with).
Moreover, this allowed us to overcome major obstacles that
 proved insurmountable for the optimization approach.
For instance,  we  have seen
examples where  goals stating that  edge lengths or
 basic part costs are all positive must be added before 
the constraints are {\prem}.  But most
 users knowing that those edges or costs are always positive, 
 (e.g., because of constraints in the database or other
rules in the program),  prefer to
submit their programs for execution  without having to
add idle conditions to make {\prem} manifest.
In these situations, the system should produce a stern warning 
message, but the user should  still be allowed to proceed with  execution.
This policy results in greater convenience  and the power of handling  situations 
that cannot be handled by optimizers.
The user can also call upon the system assistance by 
initially  executing in ``debugging mode'' under the trust-but-verify policy
outlined at the end of Section \ref{sec:aggregates}.
This will either  produce counter-examples  that show that $\gamma$ 
is not {\prem}, or after enough testing, the user  can reliably assume 
that the property holds, and the debugging mode can be turned off. 
\sinv\inv\inv\subsection*{Acknowledgements}
We would like to thank the reviewers for the comments and suggested improvements. This work was supported in part by NSF grants IIS-1218471, IIS-1302698 and CNS-1351047, and U54EB020404 awarded by NIH Big Data to Knowledge (BD2K).\inv

\inv
\bibliographystyle{acmtrans}
\bibliography{monotonic}
\appendix
\newpage
\section{Experiments}\label{sec:result}
Our experiments seek to quantify the performance improvements delivered by the constraint transfer optimization described in the previous sections,
compared with (i) the basic iterated fixpoint of stratified programs, (ii) the monotonic aggregate approach
of  \cite{mazuran2012extending},
and (iii)   an XY-stratification based approach \cite{ldl++}. In fact,
XY-stratification is of particular interest since
it  is known to be quite effective at  expressing and
implementing efficiently in Datalog complex algorithms that use aggregates in
recursion. This is because XY-stratified programs use a restricted form of 
local stratification made explicit by the 
$\tt {+}1$ form of the first argument in the
recursive predicate, whereby the iterated fixpoint
computation for basic programs becomes a basic iterative loop. In fact.
the iterative loop expressed by Example \ref{ex:sssp-xy} is 
akin  to the one of the seminaive computation of
Example \ref{ex:sssp-cpr}.

For our  experiments
we  used  the sequential version of \deals \cite{shkapsky2015optimizing}  that along
with the methods (i)--(iii),  now supports the constraint transfer optimization implemented as follows: for a given program, the system (a) parses the rules and identifies the constraints to which the optimization is applicable;
(b) generates a new set of rules through rewriting; (c) parses the rules again and produces an instantiated AND/OR tree \cite{ldl++} representing the execution plan for the rewritten rules; (d) evaluates the program using a tuple-at-a-time pipelined execution.

\stitle{Experimental Setup.} The experiments were performed on a machine with four AMD Opteron 6376 CPUs (16 cores per CPU) and 256 GB memory (configured into eight NUMA regions). The operating system is Ubuntu Linux 12.04 LTS, and the version of the JVM is Java 1.8.0\_11 from the Oracle Corporation. The \deals runtime uses only one core, but we allow the JVM to use all the available cores for background tasks like garbage collection.

\stitle{Test Programs.} We report the experimental results on using \deals to solve the single-source shortest path problem. The compared programs are listed in \ref{sec:examples}. A user can write the program in \cexp\ref{ex:sssp-stratified}, while \deals will optimize it by transferring the $\tt min$ aggregate into the recursive rules, and evaluate the program in \cexp\ref{ex:sssp-cpr}.
The non-trivial  XY-stratified Datalog program \cite{ldl++} shown in \cexp\ref{ex:sssp-xy} mimics the seminaive evaluation of the program in \cexp\ref{ex:sssp-cpr}.
Finally, the formulation that uses monotonic min $\tt mmin$ is shown in \cexp\ref{ex:sssp-fsmin}.
We denote these four different solutions to the same problem by $\tt spath$, $\tt spath\_prem$, $\tt spath\_xy$ and $\tt spath\_mmin$, respectively.

\stitle{Datasets.} We compare the performance of these four programs on synthetic graphs and real world graphs. Each $n$-node synthetic graph used in the experiments has integer node ids ranging from 0 to $n - 1$, and the length of each arc is a random integer between 1 and 100. These synthetic graphs are directed acyclic graphs (DAGs) generated by connecting each pair of vertices $i$ and $j$ ($i < j$) with probability 0.1. The real world graphs listed in Table \ref{tbl:graph} are selected from the Stanford Large Network Dataset Collection (Leskovec and Krevel 2014). These graphs do not have arc lengths, therefore we generated integers from a uniform random distribution between 1 and 100 as the values of arc lengths. For each graph, we (i) randomly select five nodes; (ii) use each node as the source node, and take the average of five runs for each program; (iii) report the min/average/max time among all five nodes in the form of error bars, while the average time is displayed above each error bar.

\begin{table}[h]
\centering
\caption{Parameters of real world graphs}\label{tbl:graph}
\begin{tabular}{lrr}
\hline\hline
Name & Nodes & Arcs \\
\hline
{\tt web-Stanford} & 282K & 2,312K \\
{\tt web-Google} & 876K & 5,105K \\
{\tt soc-Pokec} & 1,633K & 30,623K \\
{\tt soc-LiveJournal1} & 4,848K & 68,994K \\
\hline\hline
\end{tabular}
\end{table}

\begin{figure}[!t]
\centering
\includegraphics[width=0.6\textwidth]{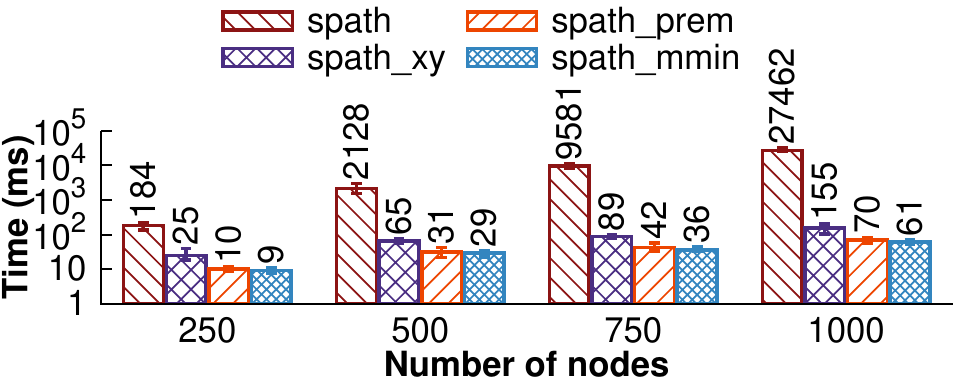}
\caption{Performance comparison on synthetic DAGs. The Y-axis is in logarithmic scale.}
\label{fig:synthetic}
\end{figure}

\stitle{Results on Synthetic Graphs.} The evaluation time for all four programs on DAGs is shown in \cfig\ref{fig:synthetic}. $\tt spath$ produces many more non-minimal tuples w.r.t. the min constraint than the other three programs. Thus, it is the slowest on all four DAGs: from 7$\times$ to 177$\times$ slower than the second slowest program. The remaining three programs produce about the same number of intermediate tuples. The performances of $\tt spath\_prem$ and $\tt spath\_mmin$ are
very close since they rely on  very similar implementations. 
However
 $\tt spath\_xy$ is about 2$\times$ slower than $\tt spath\_prem$
due to the overhead of its evaluation plan. We also compared the performance of these programs on other types of synthetic graphs that contain cycles, including random graphs and scale-free graphs. $\tt spath$ does not terminate on these graphs (more details are discussed in \ref{sec:examples}), 
while the results for the other three programs are similar to that of \cfig\ref{fig:synthetic}, and are omitted here due to space constraints.

\begin{figure}[!t]
\centering
\includegraphics[width=0.6\textwidth]{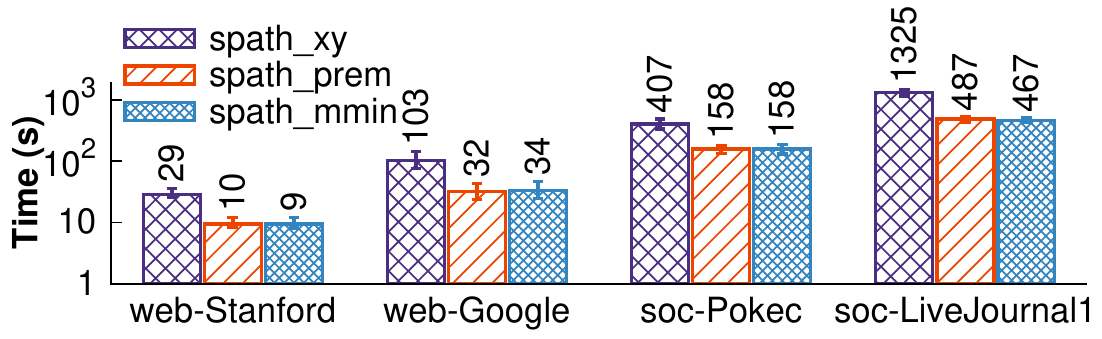}
\caption{Performance comparison on real world graphs. The Y-axis is in logarithmic scale.}
\label{fig:realworld}
\end{figure}

\stitle{Results on Real World Graphs.} We only report the results for $\tt spath\_xy$, $\tt spath\_mmin$ and $\tt spath\_prem$ in \cfig\ref{fig:realworld}, as $\tt spath$ does not terminate due to the presence of cycles in these graphs. We observe a similar trend as in the results on synthetic graphs---the time of $\tt spath\_prem$ is very close to that of $\tt spath\_mmin$, with $\tt spath\_xy$ always being the slowest, about 3$\times$ slower than the other two programs. This is again due to the more complex plan used by the XY-stratified
program that besides usability also impacts performance
on a single-processor machine. In fact we have reasons to suspect that
the performance benefits of {\prem} aggregates in recursion over aggregates in XY-stratified programs might become substantial
on parallel implementations, and this proposes an important
topic deserving further investigation. Indeed, while
scalable performance through
parallelism of programs with {\prem} aggregates was 
amply demonstrated in 
\cite{shkapsky2015optimizing,yang2015parallel}, the parallelization of XY-stratified program largely remains a topic for future research.

Overall, the optimized program obtained by transferring the aggregate into the recursive rules achieves basically the same performance as the program that uses monotonic aggregates, and much better performance than the original stratified program,  and also 
speed-ups that are twice or better than those of XY-stratified program
running on \deals\footnote{While \deals
 has been shown to deliver  better than competitive performance when compared with state-of-the-art systems such as
 DLV (Leone et al. 2006), Clingo (Gebser et al. 2014), and LogicBlox~\cite{aref2015design},
such  comparisons would take us beyond the scope
 of this paper, since the goal of our
experimental evaluation was to determine the performance of different techniques on the same problem, same system and  same runtime.}.
\inv \section{Programs Used in Experiments}\label{sec:examples}
\begin{example}[Shortest path from node $\tt a$ with stratified aggregates.]
\label{ex:sssp-stratified}
\cldl
\prule{path(Y, Dy)}{arc(a, Y, Dy).}
\prule{path(Y, Dy)}{path(X, Dx), arc(X, Y, Dxy),Dy = Dx + Dxy, Dy > Dx.}
\prule{spath(Y, \magg{Dy})}{path(Y, Dy).}
\eldl
This program is the same as the program in \cexp\ref{ex:shortestpath1}, except that we added $\tt Dy > Dx$ to the recursive rule to avoid integer overflow problems in evaluating $\tt Dy = Dx + Dxy$. The computation of
$\tt spath$ does not terminate on graphs that contain cycles, unless we enforce an upper bound constraint for  values that represent the path length.
In our experiments, we used 32-bit signed integers and this 
establishes an implicit upper bound constraint. 
But in our experiments we never reached that upper bound since we ran out of memory on our test machine that comes with a 256 GB of memory.
\end{example}

\begin{example}[Shortest path from node $\tt a$ after the transfer of the min constraint]
\label{ex:sssp-cpr}
\cldl 
\prule{path(Y, \magg{Dy})}{arc(a, Y, Dy).}
\prule{path(Y, \magg{Dy})}{path(X, Dx), arc(X, Y, Dxy),Dy = Dx + Dxy, Dy > Dx.}
\prule{spath\_prem(Y, Dy)}{path(Y, Dy).}
\eldl
\end{example}

\begin{example}[Shortest path from node $\tt a$ with XY-stratification]
\label{ex:sssp-xy}
\cldl
\prule{delta\_path(0, Y, Dy)}{\hspace{-0.5em}arc(a, Y, Dy).}
\prule{delta\_path(J + 1, Y, \magg{Dy})}{\hspace{-0.5em}delta\_path(J, X, Dx),arc(X, Y, Dxy),}
\pbody{}{\hspace{-0.5em} Dy = Dx + Dxy, Dy > Dx,}
\pbody{}{\hspace{-0.5em}if (all\_path(J, Y, D)~then~D > Dy).}
\prule{all\_path(J + 1, Y, Dy)}{\hspace{-0.5em}all\_path(J, Y, Dy),\neg delta\_path(J + 1, Y, \_).}
\prule{all\_path(J, Y, Dy)}{\hspace{-0.5em}delta\_path(J, Y, Dy).}
\prule{last\_step(J)}{\hspace{-0.5em}delta\_path(J, \_, \_),\neg delta\_path(J + 1, \_, \_).}
\prule{spath\_xy(Y, Dy)}{\hspace{-0.5em}last\_step(J), all\_path(J, Y, Dy).}
\eldl
The evaluation plan of this program is very complex---there are 120 nodes in the corresponding AND/OR tree, while both AND/OR trees of $\tt spath$ and $\tt spath\_prem$ have only 15 nodes.
\end{example}

\begin{example}[Shortest path from node $\tt a$ with monotonic aggregates]
\label{ex:sssp-fsmin}
\cldl
\prule{path(Y, mmin \langle Dy \rangle)}{arc(a, Y, Dy).}
\prule{path(Y, mmin \langle Dy \rangle)}{path(X, Dx), arc(X, Y, Dxy),Dy = Dx + Dxy, Dy > Dx.}
\prule{spath\_mmin(Y, \magg{Dy})}{path(Y, Dy).}
\eldl
This program is evaluated using {\em Monotonic Aggregate Seminaive Evaluation} (MASN)~\cite{shkapsky2015optimizing}, with an evaluation plan akin to that of \cexp\ref{ex:sssp-cpr}.
\end{example}

\section{Discussion} \label{sec:counterexamples}

In this  paper we have proposed a new  fixpoint 
approach to the semantics of recursive programs with 
aggregates, which is conducive to efficient implementation
and to a concise logic-based formulation for many
basic algorithms. However the  solution proposed here also
has some important limitations, and does not eliminate
the desirability of other past proposals including the 
 many discussed  in \cite{DBLP:journals/corr/GelfondZ14}. It is also important that
 these limitations be understood by the Datalog programmers
 and compiler writers, since the enthusiasm for Datalog in 
 BigData applications has recently lead some researchers to 
develop  parallel systems that support recursive programs with aggregates,
without providing a formal declarative semantics for program.
In the absence of a formal declarative semantics, it becomes difficult
to prove the correctness of a given implementation, and the risk that the same program might return different results on different 
platforms becomes a serious concern. 
A simple example of a program that  cannot be optimized by pushing extrema constraints is given 
below:
\begin{example}[Max constraint that cannot be pushed into recursion] \label{ex:counter}
\cldl
\pfact{p(2).}{}
\pfact{p(5).} {}
\prule{p(J1)}{p(J),  J \leq 10, J \neq 5, J1=J+2.~~~~~~~~~~~~} 
\prule{topp(J1)}{p(J1), is\_max(()(J1)).}
\eldl
\end{example}
In the iterated fixpoint computation of this program we obtain the following values  
for the argument of $\tt p$: 2 and 5 at the first step; then 2, 5, 4, at the second step;
then 2, 4, 5, 6,  at the third step, then 2, 4, 5, 6, 8,  and  2, 4, 5,  6, 8, 10  at the final step, whereby
the $\tt topp$ result is 10.

However if  we pre-apply the max to the facts and rules defining $\tt p$, we obtain
 5 at the first step and 5 again the second step, since the rule produces no  result.
 Thus the fixpoint iteration over the constrained rules  terminates by producing
 an incorrect result: i.e. the value  5 instead of 10.

While most programmers faced with the previous example will figure out that the max cannot be transferred into recursion, the question on whether the
program in Example \ref{ex:attend1} and the
following program are equivalent might not be easy for users to resolve
without the notion of pre-applicability.

\begin{example}[A simple variation of the program 
in Example \ref{ex:attend5}]
 \label{ex:attend8}.
\cldl
\prule{ attend(X)}{\hspace{-1.4cm}organizer(X).}
\prule{attend(Y)}{\hspace{-1.4cm} cntfriends(Y, N).}
\prule{cntfriends(Y, N)}{\hspace{-0.2cm}attend(X), friend(Y,X), mcount((Y),(X), N ).}
\prule{f\_attend(X)}{\hspace{-1.0cm}cntfriends(X, Nfx), Nfx \geq 3.}
\eldl
\end{example}

Now, while the original Example \ref{ex:attend1} can indeed
be obtained from the program in Example \ref{ex:attend8} above, by 
transferring $\tt  Nfx \geq$ this does not assure that the names of
friends returned by $\tt f\_attend$ above are the same as those of
$\tt attend$ in Example \ref{ex:attend5}. However the fact that the max constraint is {\prem}. suggests that equivalence
does not hold either, and simple counter-examples are in fact easy
to construct~\footnote{E.g., a clique of three friends 
where one views an organizer as a friend satisfies Example
\ref{ex:attend8} but not Example \ref{ex:attend5}.}.

The next example uses a variation from the 
example proposed by Ross and Sagiv \cite{ross1992monotonic} 
to illustrate problems and alternative solutions for
programs with aggregates.
 
\begin{example}[A program with multiple minimal models\inv]
\label{ex:ross_sagiv_example}
\cldl
\pfact{p(b).} {}
\pfact{q(b).} {}
\prule{cq(C)}  {q(X), mcount((), (X), C).}
\prule{p(a)}  {cq(C), C=1}
\prule{cp(C)}  {p(X), mcount((), (X), C).}
\prule{q(a)}  {cp(C), C=1.}
\eldl
\end{example}
This is a monotonic program, which has as least-fixpoint:
{\tt p(a), p(b), cp(1), cp(2), q(a), q(b), cq(1), cq(2)}.
However we cannot replace the monotonic counts with the 
standard counts since the transfer of max is incompatible 
with the "$= 1$" condition. If we remove the two 
"$= 1$" conditions, then the program obtained by replacing
mcount with count has a least fixpoint semantics and will
return {\tt p(a), p(b),  cp(2), q(a), q(b), cq(2)}. Finally,
if we remove {\tt C=1} from the last rule, we can replace
mcount with count in the previous, and we obtain the following model:
{\tt p(a), p(b),  cp(2), q(a), q(b), cq(1), cq(2)}.

\section{Additional References}

\bibitem[\protect\citeauthoryear{Gebser, Kaminski, Kaufmann, and Schaub}{Gebser
  et~al\mbox{.}}{2014}]{gebser2014clingo}
{\sc Gebser, M.}, {\sc Kaminski, R.}, {\sc Kaufmann, B.}, {\sc and} {\sc
  Schaub, T.} 2014.
\newblock Clingo= asp+ control: Preliminary report.
\newblock {\em arXiv preprint arXiv:1405.3694\/}.

\bibitem[\protect\citeauthoryear{Leone, Pfeifer, Faber, Eiter, ,
  et~al\mbox{.}}{Leone et~al\mbox{.}}{2006}]{leone2006dlv}
{\sc Leone, N.}, {\sc Pfeifer, G.}, {\sc Faber, W.}, {\sc Eiter, T.}, {\sc },
  {\sc et~al\mbox{.}} 2006.
\newblock The {DLV} system for knowledge representation and reasoning.
\newblock {\em TOCL\/}~{\em 7,\/}~3, 499--562.

\bibitem[\protect\citeauthoryear{Leskovec and Krevl}{Leskovec and
  Krevl}{2014}]{snapnets}
{\sc Leskovec, J.} {\sc and} {\sc Krevl, A.} 2014.
\newblock {SNAP Datasets}: {Stanford} large network dataset collection.
\newblock \url{http://snap.stanford.edu/data}.

\label{lastpage}
\end{document}